\documentclass[11pt,english,dvipsnames]{article}
\usepackage[utf8]{inputenc}
\usepackage{graphicx}
\usepackage{amsmath, amsthm}
\usepackage{babel}
\usepackage{comment,makecell}
\usepackage{float}
\usepackage{setspace}
\usepackage{subcaption}
\usepackage[margin=1.0in]{geometry}
\usepackage{lmodern} 
\usepackage{soul}
\usepackage{booktabs}
\usepackage{url}
\usepackage[comma]{natbib}
\setcitestyle{authoryear,open={(},close={)}}

\providecommand{\keywords}[1]
{
  \small	
  \textbf{\textit{Keywords---}} #1
}

\author{
  Jirka Poropudas\\[-2pt]
  {\small SportIQ} \\ {Helsinki, Finland}\\[-2pt]
  {\small \texttt{jirka.poropudas@gmail.com}}
  \and
  Topi Halme\\[-2pt]
  {\small Dept. of Information and Communications Engineering}\\[-2pt]
  {\small Aalto University, Espoo, Finland}\\[-2pt]
  {\small \texttt{topi.halme@aalto.fi}}
}

 \graphicspath{{figures/}}
 \usepackage{xcolor}
 \usepackage{comment}
 \usepackage{caption, subcaption, bm}

\newtheorem{proposition}{Proposition}
\newtheorem{corollary}{Corollary}

\newtheorem{remark}{Remark}

\def\bx{{\bm{x}}}

\DeclareMathOperator{\ORTG}{{ORTG}}
\DeclareMathOperator{\DRTG}{{DRTG}}
\DeclareMathOperator{\NetRTG}{{NetRTG}}
\DeclareMathOperator{\eFG}{{eFG\%}}
\DeclareMathOperator{\TOVpct}{{TOV\%}}
\DeclareMathOperator{\ORBpct}{{ORB\%}}
\DeclareMathOperator{\FTr}{{FTr}}
\DeclareMathOperator{\FGpct}{{FG\%}}
\DeclareMathOperator{\FTpct}{{FT\%}}
\DeclareMathOperator{\OPP}{{OPP}}

\title{\vspace{-0.5em}Decomposing Basketball Team Efficiency\\
Using the Four Factors and Shapley Values}

\date{}
\begin{document}
\maketitle

\begin{abstract}
    
Understanding how different aspects of basketball performance translate into overall team efficiency is central to analysis and strategy. This paper derives a mathematical expression that presents team offensive, defensive, and net ratings directly in terms of a team’s four factors and shooting percentages. The resulting formula captures the full nonlinear structure of how shooting, turnovers, rebounding, and free throws jointly interact to determine team efficiency. Using this relationship, we introduce a Shapley value–based framework that decomposes team net rating into a sum of contributions from individual factors. This way, different facets of the game are made commensurable. The decomposition requires only standard box-score statistics and a single parameter $\mu$, which represents the relative frequency of possession-ending free throws and remains constant for a given league and season. Consequently, it can be applied equally to individual games, full seasons, or league-wide trends. Moreover, it is immediately applicable to box-score statistics from any league or competition level by a simple change of reference point and the estimation of league- and season-specific parameter $\mu$. Empirical analyses using NBA data from 2000 to 2025 validate the offensive ratings produced by the formula against exact play-by-play data, showing a nearly perfect correlation ($r = 0.999$) and a low Mean Absolute Error of $0.188$ points per 100 possessions. These analyses further illustrate how the proposed framework can be used to quantify how teams achieve success in different ways or to explain the results of individual games.
\end{abstract}


\section{Introduction}

Basketball has been fundamentally shaped by the use of data and analytics in recent decades \citep[see, e.g.,][]{oliver:2004,kubatko:2007, terner:2021}. A fundamental question in basketball, like in other sports, is identifying and quantifying factors that contribute to success \citep{page:2007, rocha:2021}. Some of the most popular and fundamental metrics of team performance in basketball are the offensive rating ($\mbox{ORTG}$) and the defensive rating ($\mbox{DRTG}$), which measure the number of points scored or allowed per 100 possessions. Since the number of possessions is approximately equal for the two teams in a game, these efficiency ratings are strong descriptors and predictors of team success. 

To describe the different facets of offense and defense contributing to overall efficiency in more detail, \cite{oliver:2004} proposes partitioning the game into four ``independent'' dimensions: shooting efficiency, turnover rate, offensive rebounding rate, and free throw rate. These components are known as the four factors, and are widely used by analysts for evaluating how well a team performs at different parts of the game. However, despite their widespread use, an explicit mathematical relationship between the factors and a team's efficiency ratings has not been derived.

This paper fills that gap by presenting a closed-form mapping from offensive and defensive four factors to the efficiency ratings. The mapping clarifies how each component of a team’s box-score performance contributes to its overall efficiency. The dependence is non-linear and multiplicative, which aligns with common basketball intuition: to score  in a given possession, a team must \emph{both} avoid a turnover \emph{and} shoot the ball successfully. One practical implication of this is that a high turnover rate has a particularly large negative impact on the offensive rating of a team with efficient shooting and good offensive rebounding, since every turnover represents a larger missed opportunity, on average. Using the derived mapping, such relationships can now be rigorously quantified. 

The mapping also enables counterfactual analyses such as determining what a team's offensive rating \emph{would have been} had its shooting efficiency been $x\%$ instead of $y\%$, all else being equal. Deriving the mapping is nontrivial, since the efficiency ratings depend on the number of possessions, which are not immediately observed from the values of the four factors, but can be uncovered with some manipulation. Overall, the proposed methodology makes the different factors of offensive and defensive performance commensurable in a principled and mathematically justified manner.

The impact of the four factors on overall team performance has been examined in several previous studies. \cite{baghal:2012} uses structural equation modeling (SEM) to observe that the four factors are indicators of more general team traits, labeled offensive and defensive quality. \cite{cecchin:2022} considers the ratios of teams' respective offensive and defensive four factors and their effect on the teams' winning percentages using SEM. It is found that all the factors have a causal association to winning, with the effect of shooting efficiency being the largest. \cite{teramoto:2010} apply linear regression to predict team wins from the factors, and similarly conclude that the effects of shooting efficiency and turnover rate are the largest for determining offensive team success. A similar analysis is undertaken in \cite{jacobs:2017}, where it is found that the order of relative importance for the factors is shooting efficiency, turnover rate, offensive rebounding rate, and free throw rate, with weights $46\%$, $35\%$, $12\%$, and $7\%$, respectively. This approximately agrees with the weighting of $40\%$, $25\%$, $20\%$ and $15\%$ originally proposed by \cite{oliver:2004}. The impact of the four factors on match outcomes has also been studied using more complex statistical and computational models such as neural networks \citep{zimmermann:2016} and model-based recursive partitioning \citep{migliorati:2023}.

This paper proposes a novel framework for measuring team performance and allocating team success among the four factors. Using the derived closed form mapping, we do not have to resort to regressions or other latent-variable models to express team efficiency in terms of the four factors. To quantify the effect of each factor, we first define a baseline, such as the league average, and then analyze how the output of the mapping changes as the factors deviate from the baseline. Based on this analysis, observed point differentials can be cleanly decomposed and attributed to individual factors using the Shapley value method \citep{shapley:1953, lundberg:2017}. The proposed framework is flexible and can be applied to describe the outcomes of individual games, to assess how the four factors drive a team’s performance over a season, or to examine the contribution of each factor to long-run changes in league-wide offensive efficiency in the NBA.

\noindent The main contributions of this paper are:
\begin{enumerate}
\item We derive a closed-form mapping from offensive and defensive four factors (together with shooting percentages) to $\ORTG$, $\DRTG$, and $\NetRTG$.
\item We use this mapping to define a Shapley value–based framework that decomposes $\NetRTG$ into interpretable contributions from individual factors relative to a user-chosen reference point.
\item We demonstrate how the framework can be applied using only standard box-score statistics across games, seasons, and league-wide trends. The framework is applicable to many leagues and levels with a simple change of reference point.
\end{enumerate}

The rest of this paper is structured as follows. Section \ref{section:efficiency} introduces the definitions of team efficiency ratings and discusses their calculation from box score data. Section \ref{section:four_factors} introduces the four factors in more detail and presents the mappings for computing team efficiency ratings using the four factors. In Section \ref{section:shapley}, we propose using Shapley values in conjunction with the derived mapping for attributing team success to the different factors in a mathematically justified manner. Section~\ref{section:analyses} presents example analyses and case studies demonstrating how the proposed approach can be utilized. Finally, in the concluding Section \ref{section:discussion}, the practical implications of the presented findings are discussed.

\section{Methodology} \label{section:efficiency}
\subsection{Team efficiency ratings}
Traditionally, the offensive rating is defined as the number of points scored per $100$ possessions. In this paper, the multiplier ``100'' is omitted from the equations in order to avoid unnecessary clutter, i.e., $\ORTG$ refers to points scored per single ball possession 
\begin{equation}
    \ORTG = \frac{ \mbox{PTS} }{ \mbox{POSS} }.
\end{equation}

In basketball, offensive possessions, excluding those at the end of a quarter, end in one of three ways: a turnover is committed, a field goal or free throw is made, or a missed shot is rebounded by the defensive team. Of course, not all made free throws are possession-ending, as free throws typically take place in clusters of two or three. Therefore, the exact calculation of the number of possessions requires examination of play-by-play statistics detailing the events of each possession separately. To avoid sifting through the play-by-play data, one would prefer an estimate for the number of possessions that is based on the box score summary. Unfortunately, the box score provides only the total number of free throws attempted and thus the exact number of possessions can not be deduced.

\cite{oliver:2004} proposes the following approximation
\begin{equation}
    \mbox{POSS} = \mbox{FGA} + \mbox{TOV} - \mbox{ORB} + \mu\cdot\mbox{FTA}, \label{eq:poss}
\end{equation} 
where $\mbox{FGA}$, $\mbox{TOV}$, $\mbox{ORB}$, and $\mbox{FTA}$ are the team's total number of field goal attempts, turnovers, offensive rebounds and free throw attempts, respectively, and parameter $\mu$ denotes the proportion of free throws that end the possession of the ball. Possession-ending free throws are defined as those free throw attempts that can be rebounded if missed -- excluding free throws following a made field goal, viz. ``and-ones''. For alternative approximations of the number of possessions, the reader is referred to, e.g., \cite{kubatko:2007} and \cite{zimmermann:2016}.

The value of $\mu$ depends on the league and the season. \cite{kubatko:2007} uses play-by-play data from NBA seasons 2003-06 to estimate the proportion of possession-ending free throws as $\mu = 0.44$. This value is traditionally used for the NBA, but the teams' shooting profiles have changed considerably since the year 2006 and the value is no longer accurate in 2025. In the 2024-25 season, we found the proportion of possession ending free throws to be approximately $42.5\%$ (see, Appendix~A), i.e., $\mu=0.425$.\footnote{We remark that calculating possessions using Equation \eqref{eq:poss} requires that team offensive rebounds (such as rebounds that went out-of-bounds with offensive team retaining possession) are included in the count of offensive rebounds. In many leagues team rebounds are included in the box score by default. In the NBA, team offensive rebounds are usually not included in the box score summary, but can be found from many alternate sources \citep[see, e.g.][]{blackport:2024}.}

Once the relative proportion of possession-ending free throws is found, \cite{oliver:2004} calculates the offensive rating as
\begin{equation}
    \ORTG = \frac{\mbox{PTS}}{\mbox{FGA} + \mbox{TOV} - \mbox{ORB} + \mu\cdot\mbox{FTA} }. \label{eq:ortg}
\end{equation}
Analogously, the defensive rating for a team is defined as the number of points allowed ($\mbox{PTS}_{\text{OPP}}$) divided by the number of opponents' offensive possessions
\begin{equation}
    \DRTG = \frac{\mbox{PTS}_{\text{OPP}}}{\mbox{FGA}_{\text{OPP}} + \mbox{TOV}_{\text{OPP}} - \mbox{ORB}_{\text{OPP}} +\mu\cdot \mbox{FTA}_{\text{OPP}}},
\end{equation}
where the statistics for the team's opponents are calculated as total sums for its opponents over the considered time frame. The overall efficiency of a team is measured using the net rating which is the difference between the team's offensive and defensive ratings
\begin{equation}
    \NetRTG = \ORTG - \DRTG.
\end{equation}

\subsection{Computing team efficiencies from the four factors
} \label{section:four_factors}

In this section, we show that the offensive rating as defined in \eqref{eq:ortg} can equivalently be computed using the four factors. The four factors proposed by \cite{oliver:2004} are the effective field goal percentage, free throw rate, offensive rebounding rate, and turnover rate. In the following, the definitions of the factors are given as in \cite{oliver:2004}, but a useful approximation for offensive rebounding rate and an alternative scaling for turnover rate are proposed. 

Effective field goal percentage ($\eFG$) measures the success rate for field goal attempts including the extra point awarded for three-pointers:
\begin{equation}
    \eFG = \frac{\mbox{FGM}+0.5\cdot\mbox{3PM}}{\mbox{FGA}},\label{eq:efg}
\end{equation}
where $\mbox{FGM}$, $\mbox{3PM}$, and $\mbox{FGA}$ are the team's total number of field goals made, total three-point field goals made, and total field goals attempted, respectively. Free throw rate ($\FTr$) measures the team's ability to get to the free throw line and make the subsequent free throws relative to the number of field goals attempted:
\begin{equation}
    \FTr = \frac{\mbox{FTM}}{\mbox{FGA}}.\label{eq:ftr}
\end{equation}
Both metrics disregard turnovers and implicitly assume that the team has reached a shot opportunity without committing a turnover. Thus, the efficiency metrics concentrate on the shooting efficiency independent of turnover rate.

Offensive rebounding rate ($\ORBpct$) measures the offensive rebounding efficiency of a team relative to the available rebounds:
\begin{equation}
    \ORBpct = \frac{\mbox{ORB}}{\mbox{ORB} + \mbox{DRB}_{\text{OPP}}},\label{eq:orb}
\end{equation}
where $\mbox{ORB}$ is the number of offensive rebounds recovered by the team and $\mbox{DRB}_{\text{OPP}}$ denotes the number of defensive rebounds for the team's opponents.
The definition of $\ORBpct$ in \eqref{eq:orb} can be approximated as
\begin{equation}
\ORBpct = 
\frac{\mbox{ORB}}{\mbox{ORB} + \mbox{DRB}_{\text{OPP}}} \approx \frac{\mbox{ORB}}{\mbox{FGA}-\mbox{FGM} }.\label{eq:orb_approximation}
\end{equation}
This is a simplification where the denominator ignores the possibility of getting an offensive rebound from an unsuccessful final free throw. In this paper, this approximation for offensive rebounding rate is employed as teams recover missed free throws only rarely and the alternative would be the estimation of a separate offensive rebounding rate for free throws. \footnote{This potential extension is discussed in Remark \ref{rem:ORTG_alpha} in Appendix~B where it is seen that the simpler, global approximation is preferable to the more complex alternatives.}

The traditional definition of turnover rate ($\TOVpct_{\text{TRAD}}$) measures the relative frequency of turnovers compared to field goal attempts and possession-ending free throws:
\begin{equation*}
    \TOVpct_{\text{TRAD}} = \frac{\mbox{TOV}}{\mbox{FGA} + \mbox{TOV} + \mu\cdot\mbox{FTA} }, \label{eq:tov_trad}
\end{equation*}
where $\mbox{TOV}$, $\mbox{FGA}$, and $\mbox{FTA}$ are the team's total number of turnovers, field goal attempts, and free throws, respectively. In this paper, the turnover rate is defined as the portion of possessions that end in a turnover, i.e., the offensive rebounds are included in the calculation:
\begin{equation}
\TOVpct = \frac{\mbox{TOV}}{\mbox{FGA} + \mbox{TOV} - \mbox{ORB} + \mu\cdot\mbox{FTA}}.\label{eq:tov}
\end{equation}

The following result establishes a link between the four factors and a team's offensive rating.
\begin{proposition}\label{thm:ortg}
   Assume that the effect of offensive rebounds following unsuccessful possession-ending free throws is negligible, so that the offensive rebounding rate can be approximated as in \eqref{eq:orb_approximation}, and that the turnover rate is given by \eqref{eq:tov}. Then a team’s offensive rating ($\ORTG$) can be expressed as a function of its four factors, namely the effective field goal percentage ($\eFG$), free throw rate ($\FTr$), offensive rebounding rate ($\ORBpct$), and turnover rate ($\TOVpct$), as well as its field goal percentage ($\FGpct$) and free throw percentage ($\FTpct$):
\begin{equation}
    \ORTG = \frac{(1-\TOVpct)(\FTr+2\cdot\eFG)}{1-\ORBpct(1-\FGpct)+\mu\cdot\FTr/\FTpct }, \label{eq:ortg_new}
\end{equation}
where $\mu$ is the relative frequency of possession-ending of free throws. 
\end{proposition}
 
The proof of Proposition \ref{thm:ortg} is given in Appendix~B. To validate the related simplifications, a comprehensive sensitivity analysis of alternative parameter choices (including team-specific $\mu_i$ and offensive rebounds following missed possession-ending free throws) is provided in Appendix B, demonstrating that these simplifications do not introduce meaningful bias. Please note that here the offensive rating is for a single possession. If a ``per 100 possessions'' version is desired, the equation should be multiplied by one hundred.

A team's defensive and net ratings are obtained directly from Proposition \ref{thm:ortg} by observing that a team's defensive rating is equal to the offensive rating of its opponents, and the net rating is the difference between offensive and defensive ratings.
\begin{corollary}\label{cor:cor1}
    A team's defensive rating ($\DRTG$) can be calculated as
    \begin{equation}
    \DRTG = \frac{(1-\TOVpct_{\OPP})(\FTr_{\OPP}+2\cdot\eFG_{\OPP})}{1-\ORBpct_{\OPP}(1-\FGpct_{\OPP})+\mu\cdot\FTr_{\OPP}/\FTpct_{\OPP} }, \label{eq:drtg_new}
\end{equation}
and its net rating ($\NetRTG$) is given by
\begin{eqnarray}
    \NetRTG & = & \frac{(1-\TOVpct)(\FTr+2\cdot\eFG)}{1-\ORBpct(1-\FGpct)+\mu\cdot\FTr/\FTpct } \nonumber\\
    & - & \frac{(1-\TOVpct_{\OPP})(\FTr_{\OPP}+2\cdot\eFG_{\OPP})}{1-\ORBpct_{\OPP}(1-\FGpct_{\OPP})+\mu\cdot\FTr_{\OPP}/\FTpct_{\OPP} }.\label{eq:netrtg_new}
\end{eqnarray}
\end{corollary}

The most important takeaway from Proposition~\ref{thm:ortg} is the dependence between a team's four factors and its offensive rating. Originally, the four factors were intended as independent measures for four different aspects of the performance of a basketball team \citep{oliver:2004}. Proposition~\ref{thm:ortg} implies that the four factors and the two shooting percentages provide a comprehensive structural set of metrics for measuring offensive output, i.e., no additional information is needed to calculate the value of offensive rating. Furthermore, the value of the offensive rating cannot be calculated from other box score statistics without including the information encoded into the four factors making this specific combination a non-reducible set of box score statistics required to compute the efficiency. It is emphasized that the equality in \eqref{eq:ortg_new} is exact with respect to the definition of $\ORTG$ in \eqref{eq:ortg}, which is widely used and calculable directly from the box score. Therefore, the difference between the offensive rating given by Proposition~\ref{thm:ortg} and the play-by-play based $\ORTG$ stems solely from the fact that the possession approximation~\eqref{eq:poss} may not exactly match the true number of possessions. Numerical evaluations Section~\ref{section:analyses} and Appendix B indicate that the approximation is generally very accurate. Furthermore, Corollary~\ref{cor:cor1} extends this result to show that a team's net rating, i.e., the overall level of performance, can be directly derived from the combination of its offensive and defensive four factors.

Proposition~\ref{thm:ortg} further reveals the nonlinear relationship between the four factors and the offensive rating. For example, turnovers are more costly for teams that shoot efficiently because they lose more expected points per lost shot opportunity. Similarly, teams with low field goal percentage benefit more from a high offensive rebounding rate as there are more missed shots to be rebounded. This interaction between the contributions of the four factors is also the main reason why previous analyses with linear dependence between variables have produced only partial results related to the relative importance of the four factors.

Finally, Proposition~\ref{thm:ortg} allows for hypothetical analyses where some values of the four factors are given and the corresponding offensive rating is calculated. This kind of what-if analyses show how the overall performance of a team would be affected if the team could improve their factor values. In the following section, it is shown how this approach can be used together with Shapley value method to quantify the relative importance of individual factors.

\subsection{Attributing team efficiency to the four factors}\label{section:shapley}

In this section, the results of Proposition~\ref{thm:ortg} are used as basis for a principled approach for dividing the observed efficiency differences between individual factors. The discussion is streamlined by considering only the offensive rating but, based on Corollary~\ref{cor:cor1}, the analysis extends naturally to defensive rating and net rating. The proposed approach yields direct answers to questions such as ``How many points per 100 possessions did this team gain from its shooting efficiency in this game?'' or ``How much of a team's season-long net rating can be attributed to controlling turnovers?''

First, we must establish a meaningful benchmark as the impact of the four factors cannot be assessed in isolation without a baseline. Specifically, when stating, e.g., that a team gained $x$ points per 100 possessions through efficient shooting, it is necessary to clarify the reference against which this gain is measured. In the context of Proposition~\ref{thm:ortg}, this baseline is represented by a reference profile of four factor values and shooting percentages, and the choice of appropriate baseline values depends on the scope of the analysis:
\begin{enumerate}
\item League-wide trends. When studying long-term changes in league performance, a natural baseline is given by the league averages from the initial season under consideration. Comparing the following seasons to this baseline reveals how offensive performance evolves over time.
\item Team-to-team comparisons. When comparing teams within a given season, it is useful to employ the previous season’s averages as the baseline for the early rounds, since it takes a number of games for league averages to stabilize. Later in the season, the current league averages provide the most up-to-date benchmark.
\item Single-team analysis. When analyzing the performance of one team over a sequence of games, the league average offers a natural point of reference. However, depending on the team’s ambitions, benchmarking against the best teams in the league may be more informative. Conversely, if the focus is on internal development, the team’s own season-to-date averages serve as the most relevant benchmark.
\end{enumerate}

Having established a suitable baseline, the closed-form Equation~\eqref{eq:ortg_new} enables counterfactual evaluation of how a change in any of the four factors affects a team's offensive rating. For example, one way to quantify the marginal contribution of $\eFG$ with respect to a baseline would be to compare the team's actual offensive rating with the offensive rating of a team whose statistical profile is otherwise identical except for $\eFG$ which is set as equal to the reference level. The difference between the two offensive ratings is then an indication of the impact of the team's $\eFG$ with respect to the baseline value. However, since the mapping in Equation~\eqref{eq:ortg_new} is non-linear, computing such one-at-a-time comparisons for each factor individually would not yield a consistent attribution. In particular, there is no guarantee that the sum of attributions computed in this way would equal the overall difference between an observed rating and the reference rating.

In order to deal with the non-linearities and interactions present in \eqref{eq:ortg_new}, we propose that the contributions of the four factors should be measured using Shapley values \citep{shapley:1953}. The concept of Shapley values originates from cooperative game theory as a method for fairly distributing the total payoff of a coalition of players based on their marginal contributions. The unique properties of Shapley values have led to their widespread adoption in other fields, notably for explaining the output of complex machine learning models by evaluating feature importance and attributing predictions to input features \citep{lundberg:2017}. In our context, the ``cooperating players'' are the four offensive factors, and the ``total payoff'' is the change in offensive rating relative to the baseline.\footnote{One should note that because $\eFG$ and $\FGpct$ are mechanically linked, and $\FTr$ and $\FTpct$ are likewise linked, the Shapley ``players'' are defined as the four primary factors, with the respective shooting percentage changing simultaneously whenever its primary factor changes (cf. Appendix~C).}

The Shapley value $\phi_F$ of factor $F$ quantifies the factor's contribution to the difference between a team's observed offensive rating and a baseline offensive rating. This is done by evaluating the average marginal contribution of $F$ over all possible orders in which the factors can be changed from their baseline values to the observed values. For instance, to determine the contribution $\phi_{\eFG}$ of $\eFG$, we consider scenarios where the value of $\eFG$ changes first, second, third, or last relative to changes in the other three factors. To find the contribution $\phi_{\eFG}$, the marginal contribution in each $4! = 24$ possible orderings is averaged to arrive at a value that represents the contribution of $\eFG$ to the total change in the value of offensive rating. The computation of Shapley values in the considered setting is presented in Appendix~C.

The proposed Shapley values can be shown to be unique in assigning contributions to different factors such that certain natural and desirable properties are satisfied \citep{lundberg:2017}. Of these properties, the most relevant to us is \emph{efficiency} which states that the sum of the contributions attributed to each of the four factors will precisely equal the total observed change in offensive rating. That is, the Shapley values satisfy
\begin{equation}
    \ORTG - \ORTG_\text{ref} = \phi_{\eFG}+\phi_{\TOVpct}+\phi_{\ORBpct}+\phi_{\FTr},
\end{equation}
where $\ORTG_\text{ref}$ is the reference offensive rating. For a discussion on the other unique properties, see \citep{lundberg:2017}. Please note also the sign of $\phi_{\TOVpct}$ is defined so that offensive rating \emph{increases} when $\TOVpct$ \emph{decreases}.

This Shapley value framework offers a clear advantage over methods previously used for analyzing the four factors. Previously presented linear regression weights \citep{teramoto:2010, jacobs:2017} fail to capture the multi-way interactions included in Proposition~\ref{thm:ortg}, and local sensitivity measures, such as partial derivatives, violate the efficiency property and leave a portion of the rating variation unallocated. Complex interactions are captured by machine learning techniques that typically rely on opaque black-box models such as neural networks \citep[see, e.g.,][]{migliorati:2023, zimmermann:2016}. While Shapley values could be used as a variable importance metric to explain these machine learning models, the underlying predictive models remain black-box in nature. Our approach diverges by applying the Shapley framework directly to the exact mathematical formulas for the efficiency ratings. This direct application keeps the full analysis entirely transparent and mathematically tractable while guaranteeing that the allocation of credit allows for non-linear interactions and satisfies the efficiency property.

In the following section, we present examples of how Proposition~\ref{thm:ortg} can be used in conjunction with the Shapley values to decompose historical changes in league-wide offensive efficiency, analyze individual team performance relative to league average, and examine how each factor influenced the final score of any single game.

\section{Case studies and example analyses}\label{section:analyses}
This section applies the proposed framework to NBA data in three timescales: long-term league evolution, season-level team evaluation, and game-level analysis. The descriptive examples illustrate how the closed-form efficiency equations and Shapley values can be used to interpret and compare offensive performance across different time scales. Then, we use the proposed approach to evaluate the relative importance of the four factors in the NBA. Finally, the four factor analysis and Shapley decomposition are applied to three different leagues to illustrate the generalizability of the framework.

\subsection{Evaluating long term offensive trends using the Shapley decomposition}\label{subsec:long_term_NBA_analysis}

\begin{figure}
    \centering
    \includegraphics[width=0.8\textwidth]{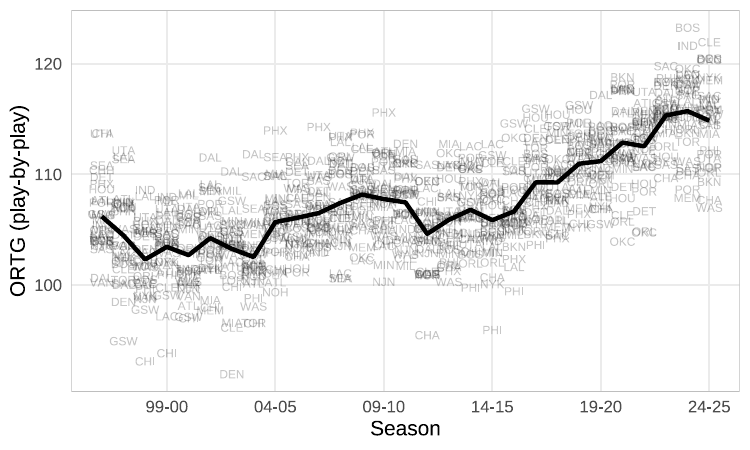}
    \caption{Evolution of offensive rating (per 100 possessions) in the NBA between seasons 1996-97 and 2024-25. The labels indicate teams and the solid line represents the league average.}
    \label{fig:ortg_evolution}
\end{figure}

\begin{figure}
    \centering
    \includegraphics[width=0.8\linewidth]{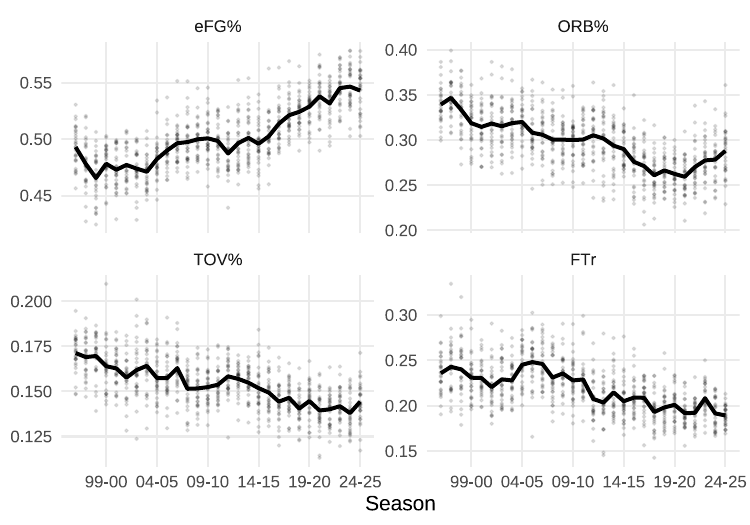}
    \caption{Evolution of the four factors in the NBA between seasons 1996-97 and 2024-25. Points indicate teams and the solid lines represent league averages. The presented values of offensive rebounding rate ($\ORBpct$) and turnover rate ($\TOVpct$) are calculated using equations \eqref{eq:orb_approximation} and \eqref{eq:tov}, respectively.}
    \label{fig:FF_evolution}
\end{figure}

The league-average offensive rating in the NBA has increased by more than 8 points per 100 possessions from the 1996-1997 season to 2024-25 (see, Figure~\ref{fig:ortg_evolution}). To better understand this development, the evolution of the values of the four factors in the NBA between seasons 1996-97 and 2024-25 is presented in Figure~\ref{fig:FF_evolution}. The league averages show how the value of $\eFG$ has increased during the time interval as the teams have emphasized three-pointers and moved away from inefficient long two-pointers. The other factors have decreased, meaning that teams have gotten to the free throw line less frequently, taken fewer offensive rebounds, and committed fewer turnovers. Based on the evolution of the four factors in Figure~\ref{fig:FF_evolution}, we can infer that the increased efficiency is broadly explained by higher league-wide $\mbox{eFG\%}$ and reduced turnover rate. On the other hand, it is also seen that offensive rebounding has been in decline until recently, which should reduce offensive efficiency.

\begin{figure}[htbp]
    \centering
    \includegraphics[width=0.8\textwidth]{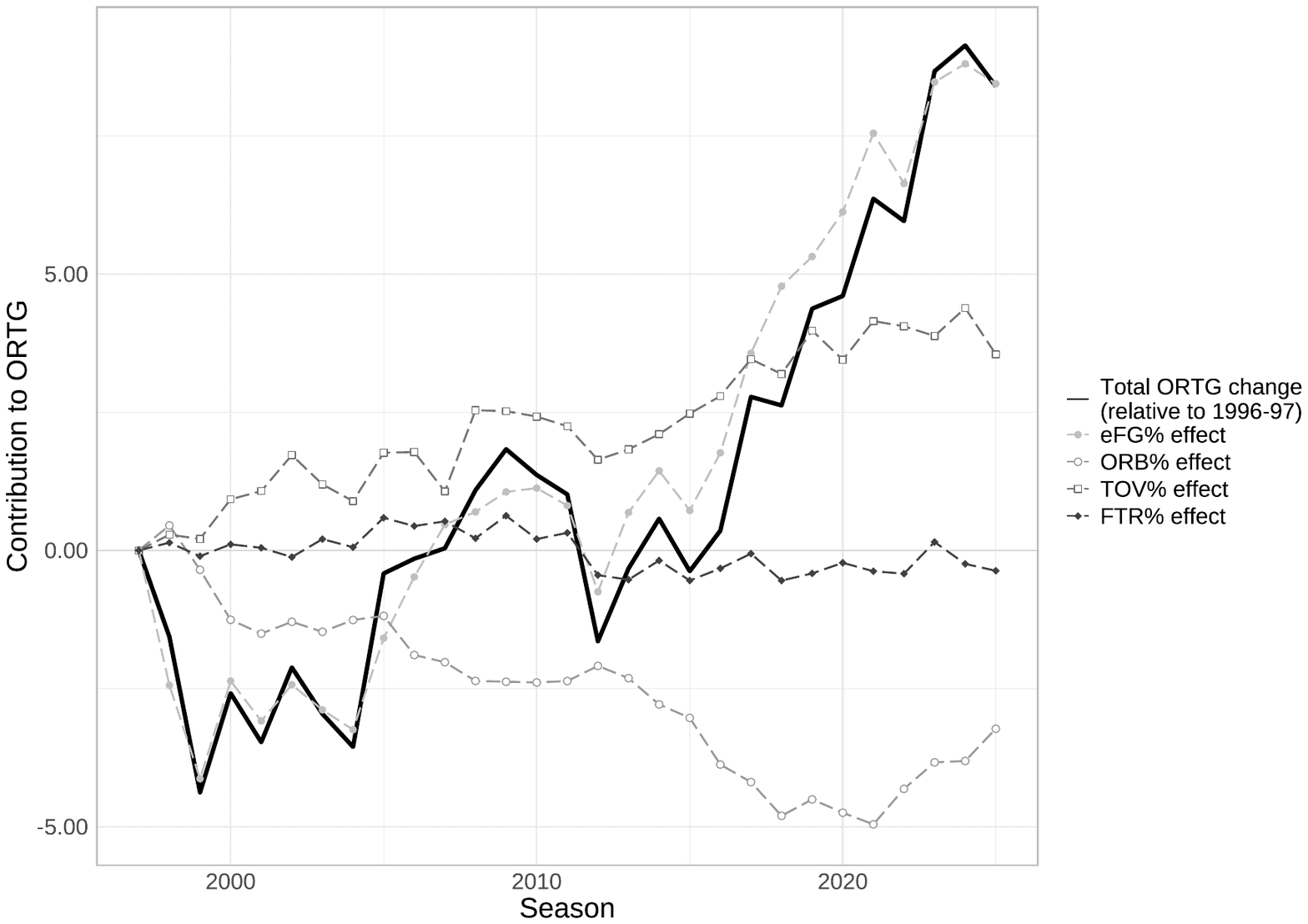}
    \caption{The Shapley contributions of different factors to the overall change in league-average offensive rating (per 100 possessions) between years 1996-25. The black line represents the change in league-wide offensive rating with respect to 1996-97 season. It is observed that the league-level evolution of offensive rating is driven mainly by $\eFG$ with impacts of $\ORBpct$ and $\TOVpct$ being almost symmetric in opposite directions.}
    \label{fig:league_evolution_shap}
\end{figure}

To accurately quantify how each of the factors has affected the evolution of the values of the offensive rating, Proposition~\ref{thm:ortg} is applied together with the Shapley method to decompose the changes in offensive rating into the contributions of the individual four factors. Figure~\ref{fig:league_evolution_shap} presents the change in league average offensive rating relative to season 1996-1997 together with the calculated contributions of the four factors. In other words, the average offensive rating and the average four factors profile for season 1996-97 are used as the baseline and other seasons are compared to that starting point. In Figure~\ref{fig:league_evolution_shap}, the total change refers to the difference between a given season's average offensive rating and the baseline, and for each season the total change equals the sum of the four factors' contributions.

The main driver of league-level change in the offensive rating is observed to be $\eFG$. As expected, the decreased $\TOVpct$ has a positive contribution on the offensive rating, while decreased $\ORBpct$ has a negative effect. Notably, the effects of these two factors are observed to be close to equal in magnitude throughout the observed time period. The contribution of changes in $\FTr$ to overall offensive efficiency is observed to be small, with an absolute contribution smaller than one point per 100 possessions.

\subsection{Team-level offensive profiles based on Shapley decomposition}\label{subsec:season_long_analysis}

While the league-wide evolution of offensive rating has been predominantly driven by improvement in effective field goal percentage, examination of individual teams reveals the wider variety of ways teams succeed offensively. The Shapley value decomposition allows to attribute a team's offensive rating deviation from the league average to each of the four factors. This section studies how teams in the 2024-25 NBA season achieved their offensive success by examining the contributions of individual factors.

\begin{table}

\caption{Raw values of the four factors and the corresponding Shapley values with respect to league average for all teams in the 2024-25 NBA season. The values of offensive rating are calculated according to Proposition~\ref{thm:ortg}. The league average offensive rating was $114.9$ points per 100 possessions.}\label{tab:team_comparison}
\centering
\begin{tabular}[!htpb]{cccccrrrrc}
\multicolumn{1}{c}{\textbf{ }} & \multicolumn{4}{c}{\textbf{Factor}} & \multicolumn{4}{c}{\textbf{Shapley value} (points per 100 poss.)} & \multicolumn{1}{c}{\textbf{ }} \\
\cmidrule(l{3pt}r{3pt}){2-5} \cmidrule(l{3pt}r{3pt}){6-9}
Team & eFG\% & ORB\% & TOV\% & FTr & eFG\% & ORB\% & TOV\% & FTr & ORTG\\
\midrule
    CLE & 0.578 & 0.295 & 0.132 & 0.187 & +5.50 & +0.45 & +1.64 & -0.11 & 122.41 \\
    BOS & 0.561 & 0.281 & 0.123 & 0.169 & +3.47 & -0.49 & +2.89 & -0.19 & 120.61 \\
    OKC & 0.560 & 0.273 & 0.117 & 0.180 & +2.60 & -1.02 & +3.67 & +0.36 & 120.54 \\
    DEN & 0.573 & 0.307 & 0.142 & 0.200 & +3.92 & +1.22 & +0.30 & +0.06 & 120.44 \\
    NYK & 0.556 & 0.299 & 0.136 & 0.186 & +1.71 & +0.68 & +1.14 & +0.22 & 118.69 \\
    MEM & 0.554 & 0.327 & 0.152 & 0.196 & +1.51 & +2.51 & -1.06 & +0.25 & 118.15 \\
    IND & 0.562 & 0.249 & 0.132 & 0.190 & +2.66 & -2.46 & +1.72 & +0.17 & 117.03 \\
    SAC & 0.548 & 0.286 & 0.135 & 0.190 & +0.47 & -0.12 & +1.31 & +0.41 & 117.00 \\
    MIN & 0.554 & 0.292 & 0.148 & 0.196 & +1.97 & +0.25 & -0.49 & +0.31 & 116.96 \\
    PHX & 0.561 & 0.257 & 0.144 & 0.195 & +2.85 & -1.99 & +0.04 & +0.61 & 116.44 \\
    LAL & 0.557 & 0.266 & 0.143 & 0.213 & +2.06 & -1.39 & +0.14 & +0.66 & 116.40 \\
    MIL & 0.568 & 0.229 & 0.135 & 0.201 & +3.89 & -3.68 & +1.31 & -0.25 & 116.21 \\
    HOU & 0.523 & 0.361 & 0.141 & 0.178 & -3.27 & +4.88 & +0.45 & -0.96 & 116.03 \\
    GSW & 0.536 & 0.310 & 0.143 & 0.187 & -0.66 & +1.41 & +0.23 & -0.33 & 115.58 \\
    DET & 0.547 & 0.301 & 0.149 & 0.192 & +0.42 & +0.82 & -0.58 & -0.03 & 115.58 \\
    LAC & 0.554 & 0.277 & 0.152 & 0.200 & +1.46 & -0.70 & -1.01 & +0.54 & 115.22 \\
    ATL & 0.546 & 0.297 & 0.151 & 0.196 & +0.32 & +0.58 & -0.90 & +0.06 & 115.00 \\
    DAL & 0.549 & 0.267 & 0.141 & 0.203 & +0.77 & -1.31 & +0.41 & +0.15 & 114.95 \\
    CHI & 0.553 & 0.255 & 0.143 & 0.173 & +1.79 & -2.14 & +0.19 & +0.02 & 114.80 \\
    SAS & 0.544 & 0.269 & 0.139 & 0.181 & +0.23 & -1.21 & +0.68 & -0.30 & 114.33 \\
    MIA & 0.544 & 0.255 & 0.140 & 0.183 & +0.17 & -2.11 & +0.63 & +0.03 & 113.66 \\
    PHI & 0.527 & 0.270 & 0.139 & 0.201 & -2.52 & -1.20 & +0.66 & +0.29 & 112.17 \\
    POR & 0.521 & 0.330 & 0.161 & 0.187 & -3.33 & +2.72 & -2.21 & -0.35 & 111.76 \\
    UTA & 0.533 & 0.309 & 0.171 & 0.195 & -1.31 & +1.36 & -3.64 & +0.05 & 111.39 \\
    NOP & 0.519 & 0.301 & 0.148 & 0.184 & -3.80 & +0.84 & -0.47 & -0.55 & 110.95 \\
    TOR & 0.522 & 0.310 & 0.153 & 0.173 & -3.39 & +1.41 & -1.19 & -0.88 & 110.89 \\
    ORL & 0.510 & 0.296 & 0.148 & 0.208 & -5.09 & +0.51 & -0.54 & +0.38 & 110.19 \\
    BKN & 0.516 & 0.289 & 0.158 & 0.189 & -3.84 & +0.08 & -1.80 & +0.09 & 109.46 \\
    CHA & 0.503 & 0.303 & 0.159 & 0.174 & -5.99 & +0.98 & -1.89 & -0.36 & 107.67 \\
    WAS & 0.512 & 0.266 & 0.154 & 0.178 & -4.73 & -1.40 & -1.31 & -0.31 & 107.18 \\
\end{tabular}
\end{table}

Table~\ref{tab:team_comparison} presents the values of the four factors, their corresponding Shapley values (relative to league average) and the offensive ratings for NBA teams in the 2024-25 regular season. For instance, the Cleveland Cavaliers (CLE) and the Oklahoma City Thunder (OKC) were two of the best offensive teams in the league with offensive ratings above 120 points per 100 possessions, more than 5 points better than the league average $114.9$. The Cavaliers' offensive strength was based almost entirely on efficient shooting, with their $\eFG$ of $57.8\%$ contributing $+5.50$ points of offensive rating above league average. The Cavaliers also received small gains from offensive rebounding ($+0.45$) and turnover rate ($+1.64$). In contrast, while OKC also benefitted from effective shooting ($+2.60$), the largest part of their offensive success is due to very low turnover rate, contributing $+3.67$ points to their offensive rating. 

As another example, the Milwaukee Bucks (MIL) and Houston Rockets (HOU) achieved almost identical offensive ratings of $116.2$ and $116.0$, slightly above league average. While the Bucks' offensive performance was largely driven by strong $\eFG$ ($+3.89$) and hindered by weak offensive rebounding ($-3.68$), the Rockets overcame their poor shooting ($-3.27$) with exceptional offensive rebounding ($+4.88$). In fact, the Rockets' almost five point gain in offensive rebounding was the second largest individual positive contribution by any factor of any team, surpassed only by the Cavaliers' $\eFG$. That is, Houston's offensive rebounding created more points than, e.g., the very high shooting efficiency of Boston Celtics or Denver Nuggets. Due to the nonlinear interactions considered by the Shapley values, in Houston's case the quite negative contribution of the $\eFG$ is accompanied by the high contribution of $\ORBpct$. This lands the team at about the average league level instead of a bottom position.

\begin{figure}
    \centering
    \includegraphics[width=0.58\linewidth]{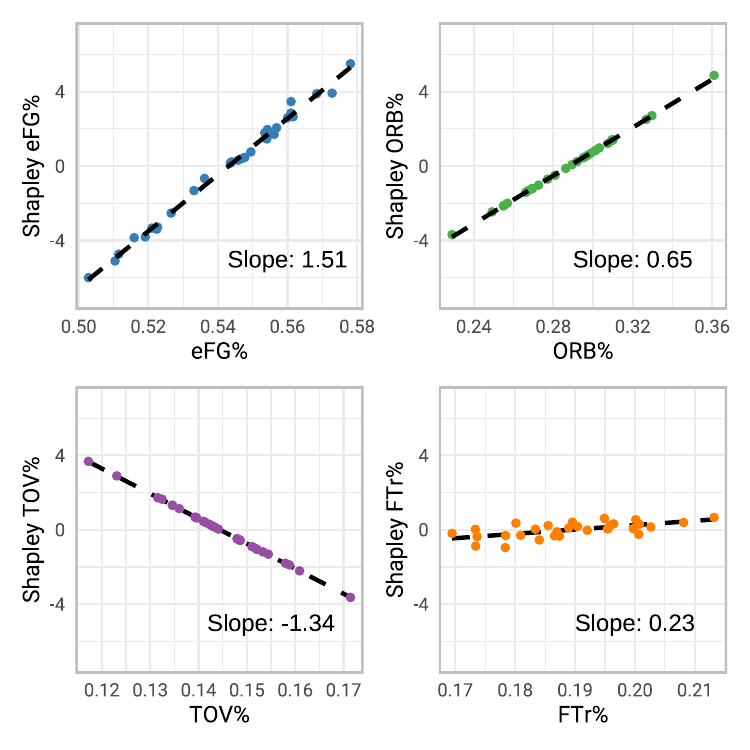}
    \caption{Relationships between raw factor values and the corresponding Shapley values for teams in the 2024-25 NBA season. For NBA teams over a full season, the relationships are near-linear within the range of observed factor values. The slopes indicate how many points of offensive rating an increase of one percentage point in a particular factor is worth for a typical team.}
    \label{fig:shapley_efg}
\end{figure}

From Table~\ref{tab:team_comparison} it is evident that effective field goal percentage is associated with Shapley values of the largest magnitude. The relationship between raw factor values and Shapley contributions is further illustrated in Figure~\ref{fig:shapley_efg}, where the factor values are plotted against their corresponding Shapley values for each factor and team in the 2024-25 season. While the mapping between the factors and the offensive rating derived in Proposition~\ref{thm:ortg} is in general highly nonlinear, it is observed that the relationship between the factor values and their corresponding Shapley contributions is mostly linear in this setting. This observed empirical linearity in the factor contributions is due to the relatively narrow spread of full-season team factor values. This suggests that within the typical range of NBA team performances, the average marginal impact of a factor scales in near-linear fashion. For example, a one percentage point increase in $\mbox{eFG\%}$ leads to an increase of approximately $+1.51$ points per 100 possessions.

\subsection{Single game analysis based on Shapley decomposition} \label{section:single_game}

Game 1 of the 2025 NBA Finals was a great example of how box score statistics can be hard to compare to each other. In this game, the Oklahoma City Thunder forced the Indiana Pacers into committing 24 turnovers, while only turning the ball over 6 times themselves. On the other hand, the Pacers shot more accurately, had more three-point attempts, and won the offensive rebounding battle 16 to 14. In the end, Tyrese Haliburton hit a buzzer beater to lift the Pacers into a 111-110 victory. The box score statistics relevant to calculating the values of four factors are presented in Table~\ref{tab:okc_ind_boxscore}. 

\begin{table}

\caption{\label{tab:okc_ind_boxscore}
Raw box score statistics from Game 1 of the 2025 NBA Finals used for calculating the four factors and the Shapley value analysis. The possession counts are estimated using Equation~\ref{eq:poss} (with $\mu = 0.425$) and the offensive ratings are calculated using Proposition~1.
}
\centering
\begin{tabular}[t]{l|cccccccc|ccc}
Team & FGM & FGA & 3PM & 3PA & FTM & FTA & OREB & TOV & POSS & PTS & ORTG\\
\midrule
IND & 39 & 82 & 18 & 39 & 15 & 21 & 16 & 24 & 98.9 & 111 & 112.2\\
OKC & 39 & 98 & 11 & 30 & 21 & 24 & 14 & 6 & 100.2 & 110 & 109.8\\
\end{tabular}
\end{table}

\begin{table}

\caption{\label{tab:g1_shapleys_multicol}
Four factors and Shapley value contributions (points per 100 possessions) for Game 1 of the 2025 NBA Finals.
}
\centering
\begin{tabular}[t]{lccccc}
 & \multicolumn{2}{c}{IND} & \multicolumn{2}{c}{OKC} & Net (IND) \\
\midrule
eFG\% & 0.585 & (+6.99) & 0.454 & (-14.59) & +21.58\\
ORB\% & 0.372 & (+5.42) & 0.237 & (-3.40) & +8.82\\
TOV\% & 0.243 & (-13.88) & 0.060 & (+10.61) & -24.50\\
FTr   & 0.183 & (-1.25)  & 0.214 & (+2.24) & -3.49\\
\hline
& & -2.72 & & -5.14& +2.42\\
\end{tabular}
\end{table}

The Pacers committed a high number of turnovers, but to which degree was it offset by their efficient shooting and offensive rebounding? In order to make the box score statistics easier to digest and compare, we converted them into values of four factors and applied the Shapley decomposition with NBA league averages for season 2024-25 as baseline (cf. Table \ref{tab:g1_shapleys_multicol}). 

The results show that the cost of Pacers' turnovers was $13.88$ points (per 100 possessions) while the team gained $6.99$ points from shooting efficiently and $5.42$ points from offensive rebounding. The Thunder gained $10.61$ points by avoiding turnovers and $2.24$ points by getting to the free throw line and making the ensuing free throws. However, the Thunder shot inefficiently from the field which cost it $14.59$ points and was unable to recover its misses which cost it $3.40$ points compared to the league average. That is, based on the Shapley values, the Thunder's poor shooting was almost as costly as the Pacers' carelessness with the ball. Overall, both the Pacers and the Thunder performed worse on offense than the league average ($-2.72$ and $-5.14$, respectively). 

The result of the game can also be examined by looking at the net difference between the contributions. For example, the Pacers gained $6.99$ and the Thunder lost $14.59$ points (per 100 possessions) in $\eFG$ compared to average, meaning that Indiana gained a total of $21.58$ points in field goal shooting over Oklahoma City. The Pacers' 18 extra turnovers had an even greater net impact of $-24.50$ points. While Indiana only had two more offensive rebounds than the Thunder, our method suggests that they gained as much as $8.82$ points over the Thunder due to rebounding. Because the Thunder shot poorly and had therefore many more opportunities for offensive rebounds, even a small margin in favor of the Pacers in total offensive rebounds implies Indiana was the better rebounding team.

\subsection{Relative importance of the four factors}\label{section:relative_importance}

Previous studies have discussed the relative importance of the four factors using linear models \citep[see, e.g.,][]{teramoto:2010,baghal:2012,jacobs:2017,cecchin:2022} or more-or-less \emph{ad hoc} weightings \citep{oliver:2004}. While the objective itself is not necessarily well-defined -- for example, the ``relative importance'' may be different for different teams based on the emphasized style of play -- in this section we show how the proposed methodology can be utilized to approach such a question. One should also note that the ``relative importance'' depends on the chosen baseline and changing the baseline fundamentally changes the analytical question being asked. In other words, the presented attributions are conditional to the baseline and not absolute intrinsic values.

The relationship between four factors and offensive rating as presented in Proposition~\ref{thm:ortg} enables new ways for analyzing the relative importance of the four factors. The analysis could be performed locally using partial derivatives of \eqref{eq:ortg_new} in the neighborhood of the league average, but the Shapley value method offers a more comprehensive alternative that takes into account the nonlinearities involved and the interactions between factors. The analyses in the previous sections have already provided indications of the relative importance of the factors. In Section~\ref{subsec:long_term_NBA_analysis}, it was observed that the magnitude of $\eFG$ impact to offensive rating is approximately double compared to $\ORBpct$ and $\TOVpct$ (cf. Figure~\ref{fig:league_evolution_shap}), while the impact of $\FTr$ is small. In Section~\ref{subsec:season_long_analysis} it was similarly noted that on average the largest Shapley value gains and losses were due to $\eFG$, followed again by $\ORBpct$ and $\TOVpct$ (see, Table~\ref{tab:team_comparison} and Figure~\ref{fig:shapley_efg}).

\begin{figure}
    \centering
    \includegraphics[width=0.8\textwidth]{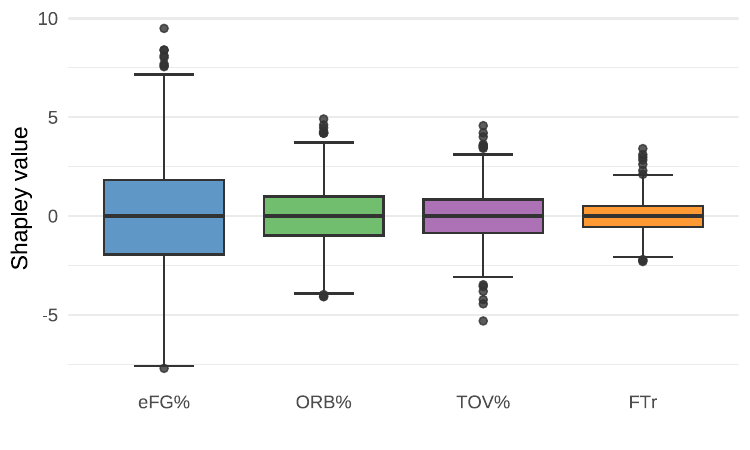}
    \caption{Box plots of season-long Shapley values for the four offensive factors across all NBA teams from 1997 to 2025. The inter-quartile ranges (IQR) illustrate the relative variability in factor contributions to offensive rating, with $\eFG$ showing the greatest dispersion and thus the largest overall impact.}
    \label{fig:shap_boxplot}
\end{figure}

Using the Shapley framework, these observations can also be quantified more systematically. The distribution of the Shapley values themselves provides a direct measure of each factor’s overall impact because the framework inherently accounts for both the sensitivity of the ORTG function and the observed variability in the values of the four factors. The more dispersed the distribution, the larger the gains or losses due to that factor. Figure~\ref{fig:shap_boxplot} presents a box plot of the team-wise Shapley gains for all seasons over the whole observation period 1997-25, where the reference point is set to the corresponding league-average of each season.
The box heights represent the inter-quartile ranges (IQR) which are, from left to right, $3.92$, $2.00$, $1.79$, and $1.10$ points per 100 possessions. When normalized to percentages, this corresponds to $45\%$, $23\%$, $20\%$ and $12\%$ which is in good agreement with the values proposed by \cite{oliver:2004}, i.e., $40\%$, $20\%$, $25\%$ and $15\%$, though our decomposition attributes relatively more importance to offensive rebounding, and relatively less to turnovers, than earlier linear estimates.

The Shapley value decomposition presented here shows that the proposed framework offers a principled and interpretable way to assess the relative impact of the four factors on offensive efficiency. While the precise numerical values depend on the league environment, i.e. the chosen baseline, the methodology provides a justified approach to a question that has traditionally been addressed through linear models or heuristic weights.

\subsection{Generalization to other basketball leagues}

The framework presented in this paper can be applied to box-score statistics from any league of competition level by simple change of reference point. To illustrate this, we compute the Shapley value contributions for all teams in the Euroleague and WNBA in the 2024-25 season. Euroleague data was obtained from \cite{basketnews2026euroleague}, and the play-by-play descriptions were retrieved using the \texttt{euroleague-api} Python package \citep{giasemidis_euroleague_api}. The WNBA data is from pbpstats.com \citep{blackport:2024}. 

Table~\ref{tab:leagues-summary} presents the average offensive rating and the values of the four factors for all three leagues. In addition, the accuracy of Proposition~\ref{thm:ortg} is evaluated by calculating the average bias, mean squared error, mean absolute error, and correlation with respect to the ``ground-truth'' season-long offensive rating for each team computed from play-by-play data. It is observed that the approximation is very accurate for all leagues. The NBA has both the most teams and most games per team, giving it the largest pooled sample for estimating the shared parameter $\mu$ and the most accurate $\ORTG$ estimates. Between the WNBA and Euroleague, Euroleague's larger number of teams (18 vs. 13) more than offsets its shorter per-team season, giving it a larger total sample and, correspondingly, better accuracy. While the WNBA bias is opposite in sign to the NBA and Euroleague, it remains very small in absolute terms ($-0.192$ points per 100 possessions, or $-0.18\%$ of the WNBA average offensive rating), well within the range expected from sampling noise given the shorter WNBA season.

\begin{table}[ht]
\centering
\caption{The summary offensive ratings (per 100 possessions) calculated using Proposition~\ref{thm:ortg} and four factors in selected leagues for season 2024-25.}
\label{tab:leagues-summary}
\begin{tabular}{lccc}
    & NBA & WNBA & Euroleague \\
    \midrule
    games & 82 & 44 & 34 \\
    teams & 30 & 13 & 18 \\
    \midrule
    ORTG & 114.9 & 104.3 & 116.1 \\
    eFG\% & 0.543 & 0.500 & 0.551 \\
    ORB\% & 0.288 & 0.288 & 0.327 \\
    TOV\% & 0.144 & 0.178 & 0.170 \\
    FTr & 0.189 & 0.213 & 0.233 \\
    FG\% & 0.467 & 0.439 & 0.476 \\
    FT\% & 0.780 & 0.787 & 0.779 \\
    $\mu$ & 0.425 & 0.434 & 0.420 \\
    \midrule
    bias & 0.076 & -0.192 & 0.119 \\
    MSE & 0.042 & 0.149 & 0.063 \\
    MAE & 0.150 & 0.317 & 0.200 \\
    correlation & 0.999 & 0.997 & 0.999 \\
\end{tabular}
\end{table}

\begin{figure}[ht]
\centering
\includegraphics[width=0.9\linewidth]{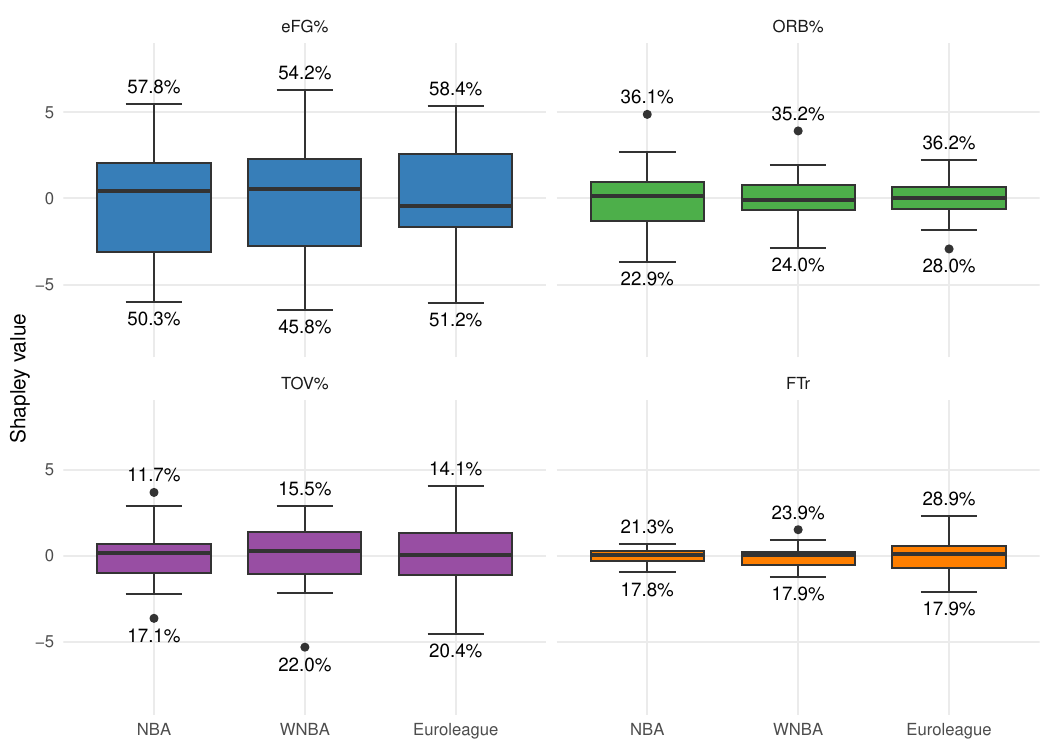}
\caption{Box plots describing the distribution of teams' Shapley values for each of the factors in the NBA, WNBA, and Euroleague in the 2024-25 season. The ranges of underlying factors are presented below and above the box plots. In all leagues, the Shapley values for $\eFG$ have the largest dispersion, indicating that effective field goal percentage is the primary driver of variation in offensive rating.}
\label{fig:all_leagues_boxplot}
\end{figure}

The league-wide four factor averages serve as the baseline reference points when evaluating team-specific performance using Shapley values. Box plots depicting the distributions of team-level Shapley values across the three leagues are shown in Figure~\ref{fig:all_leagues_boxplot}. 

To complement the visual distributions, Table~\ref{tab:interquartile_ranges} reports the inter-quartile ranges of the Shapley value contributions across each league, along with the weights relative to total variation attributed to each factor. The weights reflect the degree to which variation in a specific factor differentiates offensive efficiency within that competition. Interestingly, the NBA weights for NBA season 2024-25 differ strongly from the long-term values presented in Section \ref{section:relative_importance} due to the league average baseline that has shifted over the years.

\begin{table}[ht]
\centering
\caption{The inter-quartile ranges (IQR) and relative proportions of total Shapley value variation for selected leagues in 2024-25.}
\label{tab:interquartile_ranges}
\begin{tabular}{lccc}
    & NBA & WNBA & Euroleague \\
    \midrule
    eFG\% & 5.12 ($54\%$) & 5.08 ($52\%$) & 4.24 ($46\%$) \\
    ORB\% & 2.23 ($23\%$) & 1.47 ($15\%$) & 1.26 ($14\%$) \\
    TOV\% & 1.66 ($17\%$) & 2.48 ($25\%$) & 2.44 ($26\%$) \\
    FTr & 0.57 ($6\%$) & 0.78 ($8\%$) & 1.28 ($14\%$) \\
\end{tabular}
\end{table}

When the three leagues are compared based on Table~\ref{tab:interquartile_ranges}, the overarching structure of the Shapley value allocations remains remarkably consistent despite marked differences between league averages in Table~\ref{tab:leagues-summary}. Most importantly, effective field goal percentage exhibits the largest dispersion in all three leagues, accounting for $46\%$--$54\%$ of efficiency variation between teams. However, notable structural variations emerge when comparing factor contributions across leagues. Comparing the WNBA to the NBA, turnover rate plays a substantially larger role in differentiating team performance, representing $25\%$ of total variation ($\text{IQR} = 2.48$) compared to $17\%$ in the NBA ($\text{IQR} = 1.66$). Conversely, offensive rebounding rate contributes less to relative efficiency differences in the WNBA ($15\%$) than in the NBA ($23\%$). Similarly, in the Euroleague, both turnover rate ($26\%$) and free throw rate ($14\%$) account for larger shares of total efficiency dispersion than in the NBA ($17\%$ and $6\%$, respectively), whereas the relative contribution of $\ORBpct$ is again lower in the Euroleague ($14\%$) than in the NBA ($23\%$).

These differences reflect distinct tactical environments. In the NBA, lower average turnover rates indicate a baseline of ball security across teams, which may also reflect relatively lower defensive pressure during the regular season. On the other hand, the wide range of offensive rebounding rates allows high-volume rebounding teams to generate greater relative efficiency advantages. In contrast, higher turnover dispersion in the WNBA and Euroleague makes ball protection a more decisive differentiator. Furthermore, higher free throw frequencies in the Euroleague increase the strategic weight of free throw rate relative to the NBA.

The extreme points of the box plots highlight how these baseline differences contextualize individual team performances. For instance, as discussed in Section~\ref{subsec:season_long_analysis}, the Cleveland Cavaliers with an $\eFG$ of $0.578$ gained $5.5$ points per 100 possessions over the average NBA team. In the WNBA, the Minnesota Lynx achieved a higher gain of $6.3$ points over their league baseline despite posting a lower raw $\eFG$ of $0.542$. This greater relative contribution stems directly from the lower WNBA baseline average ($0.500$ vs. $0.543$). Similarly, Maccabi Tel-Aviv led the Euroleague with a raw $\ORBpct$ of $0.362$ but gained only $2.2$ points relative to their league average, whereas the Houston Rockets gained $4.9$ points in the NBA with a similar raw $\ORBpct$ ($0.361$), illustrating how non-linear factor interactions and distinct league baselines interact to define net performance allocations.

\section{Discussion} \label{section:discussion}

\subsection{Practical application of the four factors}

The four factors and the Shapley value-based framework provide coaches with a practical tool for monitoring team performance and development throughout a season. By establishing a baseline at the beginning of the season, coaches can evaluate how the team is progressing and identify areas that require improvement. The framework supports benchmarking and goal setting by allowing teams to define measurable objectives, such as ranking among the league leaders in points gained through offensive rebounding percentage. If the team's overall efficiency falls short of target level, the analysis helps quantify the gap and directs attention to the specific factors that require improvement.

Importantly, the framework presented in this paper identifies \emph{where} performance deficits exist, but not \emph{why} they occur. Determining the underlying causes remains the responsibility of the coaching staff. From a tactical perspective, the framework can guide game planning, opponent preparation, and resource allocation by highlighting the areas with the greatest impact on winning. It can also improve communication with players by translating statistical weaknesses into their practical consequences, such as the number of points lost per 100 possessions due to poor offensive rebounding or excessive turnovers. As illustrated in the Game 1 analysis in Section~\ref{section:single_game}, Indiana's 18 additional turnovers and Oklahoma City's inefficient shooting are not directly comparable in a standard box score, but the Shapley decomposition places both on the same points-per-100-possessions scale: Indiana's turnovers cost the team 24.50 points relative to Oklahoma City, nearly canceling out a combined 30.40-point edge in shooting efficiency and offensive rebounding. A more moderate turnover rate would have turned a game decided by a single last-second shot into a comfortable win, giving the coaching staff a quantified target for improvement rather than a general instruction to take care of the ball. 

From an execution standpoint, a weak factor value may typically reflect multiple underlying issues. For example, a low effective field goal percentage may result from poor shot creation, suboptimal shot selection, or inadequate shooting performance. Consequently, the framework can help coaches prioritize practice content and points of emphasis while recognizing that maximizing individual factor values in isolation may be counterproductive if improvements in one area negatively affect another.

The Shapley value framework also has applications in personnel management. Team- and lineup-level analyses of four factors can reveal which player combinations are most effective and help coaches understand the factors driving lineup success or failure. In addition, persistent weaknesses in specific factors may expose roster deficiencies that cannot be addressed through tactical adjustments or skill development alone. For example, a team that consistently loses the rebounding battle may benefit disproportionately from acquiring a player who provides a stronger interior presence. In this way, four factors analysis can support evidence-based decisions regarding rotations, player roles, recruitment, and player market.

\subsection{Comparison with ESPN Net Points}

One of the results of our analysis, assigning point values to contributions of individual factors, has some similarity with the Net Points framework recently introduced by \cite{oliver2025netpoints} at ESPN.
For a concrete numerical example, the analysis presented in Section~\ref{section:single_game} can be compared to ESPN Net Points. From the Pacers' perspective in Game~1 (values converted to a per-100-possessions scale), the Net Points contributions were +18.6 for field goal shooting, +7.8 for rebounds, $-20.4$ for turnovers, and $-4.7$ for free throws \citep{ESPN_net_points}. Table~\ref{tab:netpoints_comparison_g1} compares these values to the corresponding factor contributions produced by our method (Table~\ref{tab:g1_shapleys_multicol}). While the values do not match exactly, the signs and relative magnitudes are broadly consistent in this example.

\begin{table}[htpb]
\centering
\caption{Comparison of net factor contributions in Game 1 of 2025 NBA Finals from the Pacers' perspective between the proposed Shapley framework and ESPN Net Points.}
\label{tab:netpoints_comparison_g1}
\begin{tabular}{lrr}
Factor &  Shapley contributions & ESPN Net Points  \\
\midrule
eFG\% &  +21.58 &  +18.60  \\
ORB\%           &   +8.82 &   +7.80 \\
TOV\%           & -24.50 & -20.40  \\
FTr           &  -3.48 &  -4.70  \\
\midrule
Total                        &   +2.42 &   +1.30  \\
\end{tabular}
\end{table}

Empirically, we find that especially in ``typical'' games, where the factor values of the two teams are not as dissimilar as in the considered example, the factor contributions derived via our proposed methodology usually agree closely with the Net Points attributions. However, as argued in the introduction, we believe the approach proposed in this paper has merit due to its principled foundation, intuitive interpretation\footnote{As an example, the Net Points contribution for offensive turnover rate is always negative and positive for offensive free throw rate, but can be either positive or negative for offensive rebounding. As such, it is not immediately obvious what the interpretation is for a contribution equaling, say, zero. For the method proposed in this paper, a zero contribution means that a team neither lost nor gained points due to this factor, compared to the reference level.}, and applicability to many leagues and levels.

Because an objective ground-truth answer to “how much each factor contributed to the final score of a basketball game” obviously does not exist, comparing the quality of the two approaches based on results alone is difficult. We argue that the approach proposed in this paper is valuable for at least the following reasons: 1) the factor contributions are derived from first-principles by exploiting the exact dependence between the four factors and team efficiency ratings, 2) the contributions are separately interpretable for both offensive and defensive factors with respect to a user-chosen reference point, and 3) the analysis can be done using only basic box score statistics available at most leagues and levels. In contrast, the Net Points methodology is (to the best of our knowledge) proprietary, the offensive and defensive factor contributions for individual teams are not always straightforward to interpret separately, and the analysis depends on play-by-play data. Of course, the purpose of this comparison is not to criticize the Net Points approach. In fact, the inclusion of play-by-play data makes it possible to assign Net Points contributions to individual players with each factor as well as to discriminate between half-court and transition plays. Such extensions are left outside the scope of this paper.

\subsection{Future directions}

While the present findings establish a robust analytical baseline, several avenues for future research remain. The current framework does not account for home court advantage, as reflected in the home teams' net ratings \citep[see, e.g.][]{jones:2007, poropudas:2011, che:2024}, and integrating its effect into the four factors could further enhance the presented models. The remaining open questions are: ``how does the home advantage affect the four factors?'' and ``how is this effect modeled consistently with the home advantage observed in the teams' net ratings?'' On more practical level, how can the teams adjust their tactics to best exploit the home court advantage in relation to their team's strengths and weaknesses as represented by the four factors?

The Shapley value-based analysis of four factors could be extended into lineup analysis by using play-by-play data. Instead of evaluating entire teams, this granular approach would generate distinct four factor profiles for specific on-court units consisting of five players. Applying the Shapley framework at this level would isolate the precise statistical drivers of best lineups and expose the structural vulnerabilities of weaker combinations.

The current study does not explicitly model the dynamic trade-offs and dependencies between factor values. For example, a team might tactically minimize its turnover rate by limiting passing and opting for shots early in the shot clock, which inherently reduces the team's effective field goal percentage. Such trade-offs become highly relevant when optimizing endgame tactics rather than maximizing long-term averages \citep{skinner:2017}. Additionally, dependencies between offensive and defensive factors also merit further study, as lineups that excel in offensive rebounding often perform well in defensive rebounding as well. This aspect of team efficiency metrics could be explored using play-by-play data to predict the four factors for a given lineup when the individual players’ statistical profiles are known. Accounting for interactions and synergies between players could then provide a valuable tool for planning and optimizing substitution patterns.

The proposed framework can be viewed as parallel to the broader sports analytics literature that seeks to quantify the value of individual players in terms of their contribution to team performance. Traditional metrics such as value over replacement player (VORP) and wins above replacement (WAR) seek to capture how much better a player performs compared to a baseline contribution, and then translate the difference to a notion of value such as wins or win probability added. Because evaluating an individual player's impact on team performance is inherently challenging in most team sports, modern player-impact metrics are typically based on possession-level regressions that relate on-court player combinations to outcomes. This approach underlies the well-known family of adjusted plus-minus metrics in basketball \citep[e.g., SPM, RPM, RAPM, E-RAPM, ME-RAPM; see][]{engelmann:2017, grassetti:2026}, as well as recent methods proposed in other sports such as ice hockey \citep{macdonald:2012}, volleyball \citep{fellingham:2022}, and soccer \citep{matano:2023}. For a comprehensive review of plus-minus models, see \citet{hvattum:2019}.

A major frontier in sports analytics is the explicit quantification of player synergy and lineup interactions, moving beyond isolated individual evaluations. Recent methodologies utilize extended hypergraphs and simultaneous lineup-level effect modeling to show that a lineup's performance is significantly different from the mere sum of its individual parts \citep{grassetti:2021, damoulaki:2025, josephs:2026}. If taken to the lineup level, the proposed Shapley framework would offer another viewpoint to this same problem. Specifically, a single player's contribution on team efficiency could be decomposed among the four factors to see exactly how that player affects a lineup's specific tactical profile. By relying on a mapping between input and output variables, this first-principles approach would provide a clear and analytically rigorous framework for isolating individual player contributions across different lineups. 

However, applying this framework at finer resolutions introduces serious statistical constraints. As demonstrated in Appendix~B, calculating isolated parameters (such as team-specific values for $\gamma_i$ and $\varepsilon_i$) introduces notable noise due to limited sample sizes within a single season. Because lineup-level data is inherently sparser than team-level aggregates, future applications must explicitly account for sample uncertainty to maintain the integrity of the underlying mappings. Statistical resampling methods, such as bootstrapping, offer a natural mechanism to evaluate these sample size limits by exposing data insufficiency through elevated variance and sensitivity in the attributions. Resolving these empirical challenges will ultimately help answer critical questions: ``How much did an individual player contribute to the four factor profiles of the lineups they were a part of?" and ``What was the ultimate impact of this contribution on the overall efficiency of the team?''

\subsection{Conclusion}\label{section:conclusions}

This paper presents a mathematical equation that characterizes the relationship between offensive rating and the four factors of a basketball team. This equation establishes a correspondence where the four factors, when combined with field goal and free throw percentages, contain all the information necessary for calculating the offensive rating. By substituting the team's defensive four factors into the equation, it directly yields the team's defensive rating, and the difference between these ratings yields the net rating. Accordingly, the offensive and defensive four factors, along with the four shooting percentages, contain the information needed to calculate the net rating which is a commonly used measure for the overall quality of a basketball team. The resulting closed-form expression is tractable and transparent, establishing a formal foundation to analyze the joint interactions among the factors. Paired with this expression, the proposed Shapley value analysis provides a principled decomposition of team performance by making the separate facets of the game commensurable.

Building on this analytical framework, the approach proposed in this paper offers clear advantages for evaluating team performance. By converting raw box score statistics into contributions measured in points per 100 possessions, it enables the quantification of the relative contribution of individual factors. The method explicitly reveals interactions, such as those between turnovers and shooting efficiency, and clarifies how these elements interact to determine the final score. Moreover, the framework supports evidence-based decision-making in strategy development, player evaluation, and performance optimization, offering insights that are not readily apparent from traditional statistics alone.

\section*{Acknowledgments}
The authors thank the editor and the anonymous reviewers for suggestions and comments which served to improve the paper. 
The authors would also like to thank Dr. Dean Oliver and coach Tuomas Iisalo for valuable discussions.

\section*{Data Availability}
The data and code that support the findings of this study, including the scripts used to generate the figures and tables, are openly available in a public repository at \url{https://github.com/halmeto/four-factors-shapley}.

\bibliography{refs}

\section*{Appendix A} \label{section:appendix_D}

\begin{figure}
    \centering
  \begin{subfigure}{.5\textwidth}
  \centering
  \includegraphics[width=\linewidth]{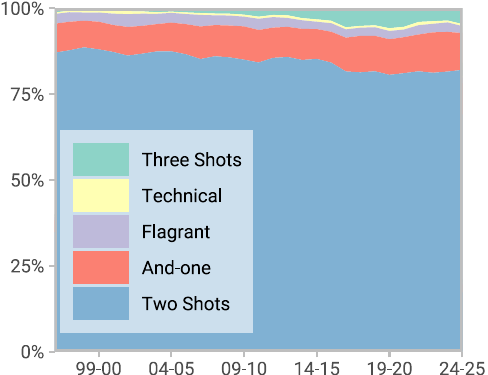}
  \caption{}
  \end{subfigure}~
 \begin{subfigure}{.5\textwidth}
  \centering
  \includegraphics[width=\linewidth]{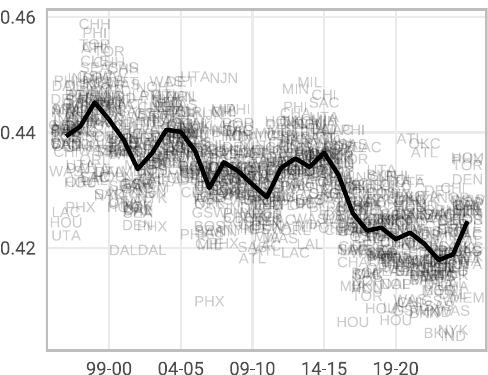}
  \caption{}
  \end{subfigure}
     \caption{(a) The distribution of free throw types in the NBA for seasons between 1996-97 through 2024-25. The proportion of possession ending free throws, $\mu$, may be obtained geometrically from the figure by adding together half of the grey area and a third of the green area. (b) The evolution of the relative frequency of possession ending free throws since 1996-1997. The solid line represents the average $\mu$ for the entire league and text labels indicate the values for individual teams.}
     \label{fig:fta_proportions}
\end{figure}

This appendix presents the updated estimate for the proportion of possession-ending free throws denoted by $\mu$ in Equation \eqref{eq:poss}.
In order to estimate the value of $\mu$ for a given league and season, the total number of free throws and the number of possession-ending free throws are calculated from play-by-play data, and $\mu$ is estimated as their ratio. The observed distribution of free throw types and the relative proportion of possession-ending free throws for NBA seasons between 1996-97 and 2024-25 are presented in Figure \ref{fig:fta_proportions}. In the figure, the ''two shots'' and ''three shots'' categories represent free throws belonging to sets of two and three shots, where only the last shot is possession-ending.  The proportion of possession-ending free throws is smaller for the latest seasons which is due to increased three-point shooting and consequently an increased number of three-shot fouls. The play-by-play data from NBA season 2024-25 suggests using $\mu = 0.425$. Moreover, it is observed that there exists some variation between teams. The effects of this variation are analyzed in Appendix B.

\section*{Appendix B} \label{section:appendix_B}

This appendix presents the proof for Proposition~\ref{thm:ortg} that shows how the offensive rating can be calculated from the four factors and shooting percentages. Additionally, some other modeling alternatives are presented and compared to the model presented in Proposition~\ref{thm:ortg}.

\subsection*{Proof of Proposition~\ref{thm:ortg}}
\begin{proof}
The points scored by the team are calculated as $\mbox{PTS} = \mbox{FTM}+2\cdot\mbox{FGM}+\mbox{3PM}$.\footnote{Three-point field goals are included in the field goal count, but produce one additional point each.} Based on the definition of the offensive rating in \eqref{eq:ortg}, the offensive rating can be written as
\begin{eqnarray}
\ORTG & = & \frac{\mbox{PTS}}{\mbox{FGA} + \mbox{TOV} - \mbox{ORB} + \mu\cdot\mbox{FTA} } \nonumber \\
& = & \frac{\mbox{FTM}+2\cdot(\mbox{FGM}+\frac{1}{2}\cdot\mbox{3PM})}{\mbox{FGA} + \mbox{TOV} - \mbox{ORB} + \mu\cdot\mbox{FTA} } \nonumber \\
& = & \frac{ \frac{\mbox{FTM}}{\mbox{FGA}} + 2\cdot\frac{\mbox{FGM} + \frac{1}{2}\cdot\mbox{3PM}}{\mbox{FGA}}}{1+\frac{\mbox{TOV}}{\mbox{FGA}}-
\frac{\mbox{ORB}}{\mbox{FGA}}+
\mu\cdot\frac{\mbox{FTA}}{\mbox{FGA}}} \nonumber \\
& = & \frac{\FTr + 2\cdot\eFG}{1+\frac{\mbox{TOV}}{\mbox{FGA}}-
\frac{\mbox{ORB}}{\mbox{FGA}}+
\mu\cdot\frac{\mbox{FTA}}{\mbox{FGA}}}, \label{eq:messedORTG}
\end{eqnarray}
where the last equality is based on the definitions of $\FTr$ and $\eFG$. An alternative representation for the denominator is obtained through the following three steps.

First, the definition of $\TOVpct$ in \eqref{eq:tov} is reorganized as
\begin{equation*}
\mbox{TOV} = \frac{\TOVpct}{1-\TOVpct}\left(\mbox{FGA}-\mbox{ORB}+
\mu\cdot\mbox{FTA}\right),
\end{equation*}
which is then divided by $\mbox{FGA}$ to yield
\begin{equation}
\frac{\mbox{TOV}}{\mbox{FGA}} = \frac{\TOVpct}{1-\TOVpct}\left(1-\frac{\mbox{ORB}}{\mbox{FGA}}+
\mu\cdot\frac{\mbox{FTA}}{\mbox{FGA}}\right). \label{eq:relativeTOV}
\end{equation}
The substitution of \eqref{eq:relativeTOV} into \eqref{eq:messedORTG} gives
\begin{eqnarray}
\ORTG & = & \frac{ \FTr + 2\cdot\eFG}{1+\frac{\TOVpct}{1-\TOVpct}\left(1-\frac{\mbox{ORB}}{\mbox{FGA}}+
\mu\cdot\frac{\mbox{FTA}}{\mbox{FGA}}\right)-
\frac{\mbox{ORB}}{\mbox{FGA}}+
\mu\cdot\frac{\mbox{FTA}}{\mbox{FGA}}} \nonumber \\
& = & \frac{ \FTr + 2\cdot\eFG}{\left(1+\frac{\TOVpct}{1-\TOVpct}\right)\left(1-\frac{\mbox{ORB}}{\mbox{FGA}}+
\mu\cdot\frac{\mbox{FTA}}{\mbox{FGA}}\right)} \nonumber \\
& = & \frac{ (1-\TOVpct)(\FTr + 2\cdot\eFG)}{1-\frac{\mbox{ORB}}{\mbox{FGA}}+
\mu\cdot\frac{\mbox{FTA}}{\mbox{FGA}}}. \label{eq:ortg_tov}
\end{eqnarray}

Second, an approximation for the term $\mbox{ORB}/\mbox{FGA}$ is construed. The approximation for $\ORBpct$ presented in \eqref{eq:orb_approximation} is equivalent to $\mbox{ORB}\approx\ORBpct\cdot(\mbox{FGA}-\mbox{FGM})$. When the approximation for $\mbox{ORB}$ is divided by $\mbox{FGA}$ one gets
\begin{eqnarray}
\frac{\mbox{ORB}}{\mbox{FGA}} & \approx & \frac{\ORBpct\cdot (\mbox{FGA}-\mbox{FGM} )}{\mbox{FGA}} = \ORBpct\cdot \left( 1 - \FGpct \right), \label{eq:relativeORB}
\end{eqnarray}
where the field goal percentage is defined as $\FGpct=\mbox{FGM}/\mbox{FGA}$.

Third, the final term in the denominator of \eqref{eq:ortg_tov} is expanded to reveal the definition of $\FTr$.
\begin{equation}
\frac{\mbox{FTA}}{\mbox{FGA}} = \frac{\mbox{FTA}}{\mbox{FTM}}\cdot\frac{\mbox{FTM}}{\mbox{FGA}}= \frac{\FTr}{\FTpct}. \label{eq:relativeFTA}
\end{equation}
Substituting \eqref{eq:relativeORB} and \eqref{eq:relativeFTA} into \eqref{eq:ortg_tov} yields
\begin{eqnarray}
\ORTG 
& = & \frac{ (1-\TOVpct)(\FTr + 2\cdot\eFG)}{1-\frac{\mbox{ORB}}{\mbox{FGA}}+
\mu\cdot\frac{\mbox{FTA}}{\mbox{FGA}}}\nonumber\\
& = & \frac{(1-\TOVpct)(\FTr+2\cdot\eFG)}{1-\ORBpct\cdot \left( 1 - \FGpct \right) +\mu\cdot\FTr/\FTpct }, \label{eq:ortg_app}
\end{eqnarray}
which completes the proof.
\end{proof}

\begin{remark}\label{rem:ORTG_mu_i}
Proposition \ref{thm:ortg} assumes that the value of $\mu$ is constant for all teams for a given league and a given season. In reality, the teams have different shot profiles which can affect the proportion of possession-ending free throws.  This detail can be included in Equation \eqref{eq:ortg_app} by estimating the value of $\mu_i$ from play-by-play data separately for each team $i$.
\begin{equation}
\ORTG 
= \frac{(1-\TOVpct)(\FTr+2\cdot\eFG)}{1-\ORBpct\cdot \left( 1 - \FGpct \right) + \mu_i \cdot\FTr/\FTpct }. \label{eq:ORTG_mu_i}
\end{equation}
\end{remark}

\begin{remark}\label{rem:ORTG_alpha}
Proposition \ref{thm:ortg} could also be expanded by considering offensive rebounds separately for missed field goals and missed possession-ending free throws. Let us denote the probability of offensive team rebounding a possession-ending free throw as $\varepsilon = \alpha\cdot\left(1-\beta\right)$, where $\alpha$ is the probability of getting an offensive rebound from a missed free throw and $\beta$ is the probability of making a possession-ending free throw. Additionally, let us denote the team-specific probability of offensive team $i$ rebounding a missed field goal as $\gamma_i$. The total number of offensive rebounds for team $i$ can then be approximated with $\mbox{ORB} \approx \gamma_i \cdot (\mbox{FGA}-\mbox{FGM} )+\varepsilon\cdot\mu\cdot\mbox{FTA}$. Substituting this approximation into Equation \eqref{eq:ortg_tov} leads to
\begin{equation}
\ORTG 
= \frac{(1-\TOVpct)(\FTr+2\cdot\eFG)}{1-\gamma_i \cdot \left( 1 - \FGpct \right) +\mu \cdot \left( 1 - \varepsilon \right)\cdot\FTr/\FTpct }, \label{eq:ORTG_alpha}
\end{equation}
where $\gamma_i$ replaces $\ORBpct$ to avoid double-counting the missed free throws rebounded by the offensive team, and the multiplier $\left(1-\varepsilon\right)$ is a correction that adjusts the proportion of possession-ending free throws $\mu$ for offensive rebounds.
\end{remark}

\subsection*{Model comparison}

In the following numerical example, the models described in Proposition~1 as well as in Remarks~\ref{rem:ORTG_mu_i} and \ref{rem:ORTG_alpha} are compared using actual NBA data. Table~\ref{tab:model_comparison} presents the accuracy of the offensive ratings calculated from four factors by comparing them with the offensive ratings calculated from play-by-play data for each of the teams and for all NBA seasons from 2000-01 to 2024-25 \citep{nba:2024,blackport:2024}. The table includes also a hybrid model that combines Remarks 1 and 2. In this model, $\mu_i$, $\gamma_i$, and $\varepsilon_i$ are estimated separately for each team. 

\begin{table}
    \centering
    \caption{The summary of different models and offensive ratings calculated from four factors for all NBA teams in seasons 2000-01 through 2024-25. The exact offensive rating (per 100 possessions) was calculated from play-by-play data and is used as baseline. Please note that in first two models $\ORBpct$ covers also the missed free throws rebounded by the offensive team (*).}
    \begin{tabular}{ccccc}
             & Proposition 1 & Remark 1 & Remark 2 & Remarks 1+2 \\ 
    \midrule
    possession-ending FTs     &  $\mu$ &  $\mu_i$ & $\mu$ & $\mu_i$ \\
    ORB from missed FGs     & $\ORBpct$ & $\ORBpct$ & $\gamma_i$ & $\gamma_i$ \\
    ORB from missed FTs  & (*) & (*) & $\varepsilon_i$ & $\varepsilon_i$ \\ 
    \midrule         
    bias     & 0.048 & 0.048 & 0.132  & 0.132  \\
    relative bias & 0.045\% & 0.044\% & 0.122\% & 0.122\% \\
    MSE     & 0.057 & 0.040 &  0.076 &  0.060\\
    MAE     & 0.188 & 0.160  & 0.221  & 0.195 \\
    correlation     & 0.999  &  0.999 & 0.999  & 0.999 \\
    \end{tabular}
    \label{tab:model_comparison}
\end{table}

The model based on Remark 1 is found to be the most accurate with a bias that is only $0.044\%$ of the average offensive rating ($108.1$ points per 100 possessions) and the smallest MSE. The model based on Proposition~\ref{thm:ortg} has equal bias and only slightly larger MSE. The models that are based on Remark 2 are less accurate compared to the first two as the more detailed expression for offensive rebounds only adds bias and noise to the results. This makes sense as the possession-ending free throws are rarely rebounded by the offensive team, i.e., the probability $\varepsilon$ is small and, on average, players have a higher probability of making the final free throw compared to other free throws, i.e., $\beta>\FTpct$ \citep{morgulev:2019}. For example in the NBA season 2024-25, the league averages were $\alpha = 0.131$, $\beta = 0.795$, and $\varepsilon = 0.026$. Furthermore, the values of $\alpha$ and $\beta$ are not included in the box score and necessitate going through the play-by-play data.

\begin{figure}[htpb]
    \centering
    \includegraphics[width=0.65\linewidth]{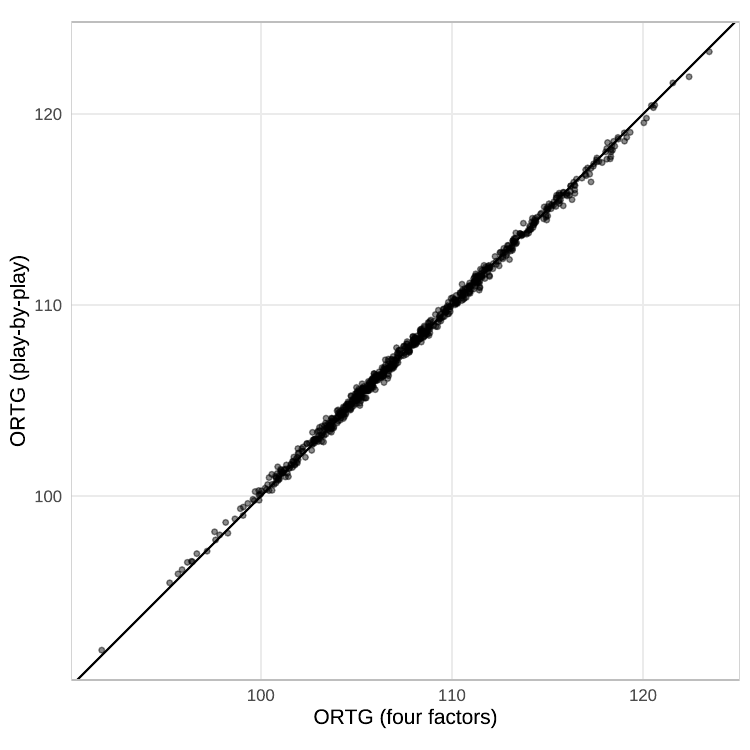}
    \caption{Offensive ratings calculated from four factors using Proposition \ref{thm:ortg} in comparison with the values calculated from play-by-play data for the NBA teams in seasons 2000-01 through 2024-25. The solid line indicates a theoretical perfect correspondence. The coefficient of correlation between values is $0.999$.}
    \label{fig:ortg_fit}
\end{figure}

Based on this comparison, the model presented in Proposition \ref{thm:ortg} is recommended over the more detailed alternatives. The model based on Remark 1 may have slightly smaller MSE, but it requires the estimation of a separate $\mu_i$ estimate for each team and season from play-by-play data. Figure \ref{fig:ortg_fit} illustrates the accuracy of the values of the offensive rating calculated using Proposition \ref{thm:ortg} by comparing them with the season-long offensive ratings of each team based on play-by-play data from NBA seasons 2000-01 through 2024-25 \citep{nba:2024,blackport:2024}. In the figure, the solid line indicates a theoretical perfect correspondence and it is seen that the fit is very good and there is no significant systematic bias.

\section*{Appendix C} %

\def\calS{{\mathcal{S}}}
\def\calF{{\mathcal{F}}}

This appendix presents the practical implementation of the Shapley value method in the context of this paper. As presented in Proposition~\ref{thm:ortg} the offensive rating of a team can be calculated from the four factors using Equation~\eqref{eq:ortg_new}. Let us now denote a four factor profile including the relevant shooting percentages as $\bx = (x_{\eFG}, x_{\TOVpct}, x_{\ORBpct}, x_{\FTr}, x_{\FGpct}, x_{\FTpct})$ and the mapping between factors and offensive rating as $g(\bx)$. Then,
\begin{equation}
    \ORTG = g(\bx)= \dfrac{(1-x_{\TOVpct})(x_{\FTr} + 2x_{\eFG})}{1-x_{\ORBpct}(1-x_{\FGpct})+\mu \cdot x_{\FTr}/x_{\FTpct}}.
\end{equation}
Let us denote the four factor profile for a given team ($T$) of interest by 
\begin{equation}
\bx^{(T)} = (x_{\eFG}^{(T)}, x_{\TOVpct}^{(T)}, x_{\ORBpct}^{(T)}, x_{\FTr}^{(T)}, x_{\FGpct}^{(T)}, x_{\FTpct}^{(T)}).
\end{equation}
For comparing the effects of individual factors we need a baseline such as the league average, denoted by
\begin{equation}
    \bx^{(0)} = (x_{\eFG}^{(0)}, x_{\TOVpct}^{(0)}, x_{\ORBpct}^{(0)}, x_{\FTr}^{(0)}, x_{\FGpct}^{(0)}, x_{\FTpct}^{(0)}).
\end{equation}
The difference between the team's offensive rating and the baseline is $\Delta(\bx^{(T)}, \bx^{(0)})=g(\bx^{(T)}) - g(\bx^{(0)})$. Now, we utilize Shapley values to divide this difference between the contributions of individual factors.

In the Shapley value method, each factor has two possible values, i.e., one observed for the team of interest and one representing the baseline. Let $\calS = \{\eFG, \TOVpct, \ORBpct, \FTr\}$ be a set consisting of the four factors. For a subset $\calF \subseteq \calS$, we denote by $\bx^{(\calF)}$ the factor vector when factors in $\calF$ take their observed values, while factors not in $\calF$ take their baseline values. For example, if $\calF = \{\TOVpct\}$, 
\begin{equation}
    \bx^{(\calF)} = (x_{\eFG}^{(0)}, x_{\TOVpct}^{(T)}, x_{\ORBpct}^{(0)}, x_{\FTr}^{(0)}, x_{\FGpct}^{(0)}, x_{\FTpct}^{(0)}),
\end{equation}
which represents a hypothetical performance otherwise at the baseline level, but with the value of $\TOVpct$ equal to the value observed for the team in question. 

It is important to note that changes in $\eFG$ are intrinsically linked to changes in $\FGpct$, and changes in $\FTr$ are linked to changes in $\FTpct$, because they measure partially the same aspects of team performance. In other words, the effects of primary factors ($\eFG$ and $\FTr$) and dependent factors ($\FGpct$ and $\FTpct$, respectively) can not be analyzed separately. In the following, the values of the dependent factors are set according to whether their primary linked factor is at its baseline or target value. That is, if $\eFG \in \calF$, $x_{\eFG}^{(T)}$ and $x_{\FGpct}^{(T)}$ are included in $\bx^{(\calF)}$, and if $\FTr \in \calF$, both $x_{\FTr}^{(T)}$ and $x_{\FTpct}^{(T)}$ are used. For instance, if $\calF = \{\eFG,\ORBpct\},$
\begin{equation}
    \bx^{(\calF)} = (x_{\eFG}^{(T)}, x_{\TOVpct}^{(0)}, x_{\ORBpct}^{(T)}, x_{\FTr}^{(0)}, x_{\FGpct}^{(T)}, x_{\FTpct}^{(0)}),
\end{equation}
and if $\tilde{\calF} = \{\TOVpct,\ORBpct,\FTr\}$
\begin{equation}
    \bx^{(\tilde{\calF})} = (x_{\eFG}^{(0)}, x_{\TOVpct}^{(T)}, x_{\ORBpct}^{(T)}, x_{\FTr}^{(T)}, x_{\FGpct}^{(0)}, x_{\FTpct}^{(T)}).
\end{equation}

The gain over the baseline offensive rating received by having factors in $\calF$ replaced by the team's actual observed values is then given by
\begin{equation}
    v(\calF) = g(\bx^{(\calF)})- g(\bx^{(0)}).
\end{equation}
Observe that $v(\emptyset) = 0$ and $v(\calS) = g(\bx^{(T)})-g(\bx^{(0)}) = \Delta(\bx^{(T)}, \bx^{(0)})$. 

The Shapley value for factor $F \in \calS$ is denoted by $\phi_F$, and is given by the equation \citep{shapley:1953,lundberg:2017}
\begin{align}
    \phi_F &= \sum_{\calF\subseteq\calS\setminus\{F\}}\dfrac{|\calF|!(|\calS|-|\calF|-1)!}{|\calS|!}\left[v(\calF\cup\{F\})-v(\calF)\right] \\
    &=\frac{1}{|S|}\sum_{\calF\subseteq\calS\setminus\{F\}}\binom{|S|-1}{|\calF|}^{-1}\left[g\left(x^{(\calF\cup \{F\})}\right) - g\left(x^{(\calF)}\right)\right].
\end{align}
where $|\calF|$ denotes the cardinality of $\calF$, and $|\calS| = 4$.
For a subset $\calF \subseteq \calS\setminus \{F\}$, the marginal contribution obtained by including factor $F$ to $\calF$ is given by $v(\calF\cup\{F\})-v(\calF) = g(x^{(\calF\cup \{F\})}) - g(x^{(\calF)})$. In other words, $v(\calF\cup\{F\})-v(\calF)$ represents the gain in offensive rating a team would receive if its factor $F$ was $x_F^{(T)}$ instead of $x_F^{(0)}$, when the other factors are given by $\bx^{(\calF)}$. The sum is taken over all subsets of the factors that do not contain $F$. Each term is weighted by the probability that, for a randomly chosen ordering of the factors transitioning from $\bx^{(0)}$ to $\bx^{(T)}$, the set of factors that have transitioned to their target values before $F$ is exactly $\calF$.

The efficiency property of Shapley values guarantees that the difference $\Delta$ between the team's offensive rating and the reference offensive rating is distributed to the four factors in its totality, i.e.
\begin{equation}
    \Delta = g(\bx^{(T)}) - g(\bx^{(0)}) = \phi_{\eFG}+\phi_{\TOVpct}+\phi_{\ORBpct}+\phi_{\FTr}.
\end{equation}
Shapley values for the defensive four factors are calculated analogously. Finally, the effect on net rating is calculated for offensive and defensive four factors separately, see, Equation \eqref{eq:netrtg_new}.

\end{document}